\documentclass[a4paper,11pt,numreference,epsfig,cite]{article}

\usepackage[total={4.5in, 9.6in}]{geometry}
\usepackage[latin1]{inputenc}
\usepackage{textcomp}
\usepackage{graphicx}
\usepackage{xcolor}
\usepackage{amsmath}

\newcommand{\stt}{\small\tt}

\newcommand{\Abstract}[1]{\begin{center}{\stt ABSTRACT}\end{center}{#1}
\vskip 0.4in}

\def\Journal#1&#2&#3(#4){\unskip, #1~{\bf #2} (#4) #3}

\def\PTP{Prog.\ Theor.\ Phys.}
\def\PTEP{Prog.\ Theor.\ Exp.\ Phys.}

\def\PRL{Phys.\ Rev.\ Lett.}
\def\PRD{Phys.\ Rev.\ D}

\def\etal{{\it et al.}}

\def\U4S{\Upsilon (\mbox{4S})}

\def\b0b0{${\rm B^{o}\overline{B^o}}$}

\newcommand{\be}{\begin{equation}}
\newcommand{\ee}{\end{equation}}
\newcommand{\ba}{\begin{array}{c}}
\newcommand{\ea}{\end{array}}
\newcommand{\beqn}{\begin{eqnarray}}
\newcommand{\eeqn}{\end{eqnarray}}

\title{Measuring the angle $\alpha_{ds}$ of the flattest Unitary Triangle with $\overline{B}_{d}\to \phi \overline{K}^{(*)0},\overline{B}_{s}\to \phi{K}^{(*)0}$ decays\protect\\}

\author{R. Aleksan$^1$, L. Oliver$^2$ , E. Perez$^3$\\
\footnotesize $^1$IRFU, CEA, Universit\'e Paris-Saclay, 91191 Gif-sur-Yvette cedex, France \\
\footnotesize $^2$IJCLab, P\^ole Th\'eorie, CNRS/IN2P3 et Universit\'e Paris-Saclay, \\ 
\footnotesize B\^at. 210, 91405 0rsay cedex, France \\
\footnotesize $^3$CERN, EP Department, Geneva, Switzerland }

\begin{document}

\maketitle

\Abstract{\normalsize \baselineskip 22pt
We show that the angle $\alpha_{ds}$ of the ``flattest'' unitarity triange can be directly measured using the decays $\overline{B}_{d}\to \phi \overline{K}^{(*)0}$ and $\overline{B}_{s}\to \phi{K}^{(*)0}$. Using both  $\overline{B}_{d}$ and  $\overline{B}_{s}$ enables a further consistency test since the expected time-dependent CP violating  asymmetries are identical though with opposite signs. Since large statistics of $\overline{B}_{d}$ and  $\overline{B}_{s}$ are needed for accurate measurements, FCC-ee and its environment at the Z-pole is well suited for such studies. These measurements, the precision of which could reach the sub-degree level, will contribute to probe further the consistency of the CP sector of the Standard Model with unprecedented level of accuracy. The main detector requirements that are set by these measurements are also outlined.}


\section{Introduction}
    
\noindent The very high statistics anticipated at FCC-ee~\cite{fccee:1,fccee:2,fccee:3} open new possibilities for studying Flavor Physics and CP violation. An endeavor that can be taken over with the FCC statistics at the Z-pole, where more that $5\cdot 10^{12}$ Z bosons should be accumulated, would be to probe with an unprecendented accuracy the CP sector of the Standard Model (SM) and to measure directly as many angles of the CKM unitary triangles. In recent papers, we have proposed to measure directly the 3 angles of a flat unitarity triangle~\cite{AOP:1,AOP:2}. 
In the present paper we propose to measure one of the angles of the flattest unitary triangle. 

\section{ Definition of the Unitary Angles}

\noindent In the SM, one derives the unitarity relations from the CKM quark mixing matrix~\cite{CKM:1},
\be
V_{CKM} =
\begin{bmatrix}
V_{ud} & V_{us} & V_{ub} \\
V_{cd} & V_{cs} & V_{cb} \\
V_{td} & V_{ts} & V_{tb} \\
\end{bmatrix}
\label{eq:CKMmatrix}
\ee

\begin{figure}[hbt]
\vfill
\begin{center}

\includegraphics[width=0.75\textwidth]{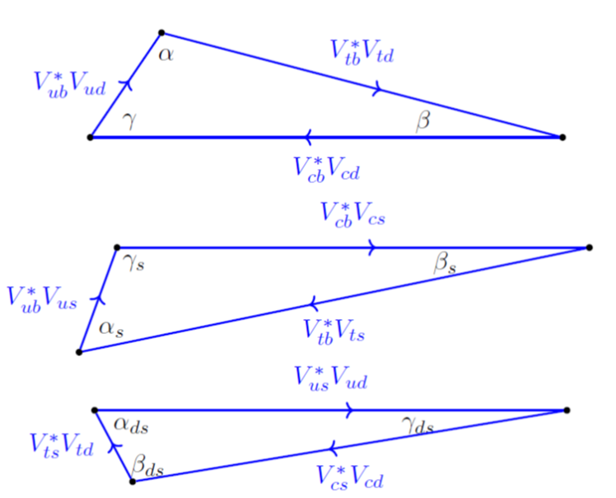}

\caption{\footnotesize \label{fig:UT654} Unitarity Triangle UT$_{db}$ involving the $1^{st}$ and $3^{rd}$ columns (top),  Unitarity Triangle UT$_{sb}$ involving the $2^{nd}$ and $3^{rd}$ columns (center) and Unitarity Triangle UT$_{ds}$ involving the $1^{st}$ and $2^{nd}$ columns (bottom) of the CKM matrix. Note that these triangles are not to scale.}
\end{center}
\vfill

\end{figure}

\noindent Should there be only 3 families of quarks, the unitarity relations, which are derived from $V_{CKM}V_{CKM}^{-1}=1$, read as below:

 \be
\begin{array}{cccccl} 
UT_{db} & \equiv & V_{ub}^* V_{ud}  + V_{cb}^* V_{cd}  + V_{tb}^* V_{td} & = & 0 \\
UT_{sb} & \equiv & V_{ub}^* V_{us}  + V_{cb}^* V_{cs}  + V_{tb}^* V_{ts} & = & 0 \\
UT_{ds} & \equiv & V_{us}^* V_{ud}  + V_{cs}^* V_{cd}  + V_{ts}^* V_{td} & = & 0 \\
\end{array}
\label{eq:UT}
\ee
In Equation~\ref{eq:UT}, only 3 relations have been displayed and are visualized in Figure~\ref{fig:UT654}. There are also 3 additional ones, but they are very similar to those above. In the SM, the CKM matrix has only 4 independent parameters. Therefore the angles of all these triangles can be expressed in terms of 4 angles~\cite{akl:1}. The first relation in Equation~\ref{eq:UT} is known as the Unitarity Triangle, with the 3 sides of the same order, and has been studied extensively. However the other ones would deserve to be studied in detail as well in order to investigate further the consistency of the SM. 
\noindent We define the angles of these triangles as 
\be
\begin{array}{cccccl} 
\alpha =  \arg \left(-\frac{  V_{tb}^* V_{td} }{ V_{ub}^* V_{ud} }\right)    ,   
\beta =   \arg \left(-{ V_{cb}^* V_{cd} \over V_{tb}^* V_{td} }\right) , 
\gamma =  \arg \left(-{  V_{ub}^* V_{ud} \over V_{cb}^* V_{cd} }\right)  \\
 \end{array}
\label{eq:UT6-angles}
\ee

\be
\begin{array}{cccccl} 
\alpha_s =  \arg \left(-\frac{  V_{ub}^* V_{us} }{ V_{tb}^* V_{ts} }\right)    ,   
\beta_s =   \arg \left(-{ V_{tb}^* V_{ts} \over V_{cb}^* V_{cs} }\right) , 
\gamma_s =  \arg \left(-{  V_{cb}^* V_{cs} \over V_{ub}^* V_{us} }\right)  \\
\end{array}
\label{eq:UT5-angles}
\ee

\be
\begin{array}{cccccl} 
\alpha_{ds} =  \arg \left(-\frac{  V_{us}^* V_{ud} }{ V_{ts}^* V_{td} }\right)    ,   
\beta_{ds}  =  \arg \left(-{  V_{ts}^* V_{td} \over V_{cs}^* V_{cd} }\right)  , 
\gamma_{ds} =   \arg \left(-{ V_{cs}^* V_{cd} \over V_{us}^* V_{ud} }\right) \\
\end{array}
\label{eq:UT4-angles}
\ee
For the angles $\alpha , \ \beta , \ \gamma$ of the triangle UT$_{db}$ we have adopted the usual convention with circular permutation of the quarks $t, \ c, \ u$. 
For the triangle UT$_{sb}$, we have adopted the generally accepted notation for $\beta_s$ in~\cite{pdg:1} and for the other angles the corresponding circular permutations of $t, \ c, \ u$. For the triangle UT$_{ds}$, we have used the notation of UT$_{sb}$ with the replacement $b \to s, \ s\to d$. This latter triangle is the flattest one since the angle $\gamma_{ds}$ is almost $\pi$. Measuring directly this angle is very difficult, however it is possible, as we will show here, to measure $\alpha_{ds}$ with some specific B decays.

\section{$B_{d,s}\to \phi K^{*0} ( \overline{K}^{*0})$ or $B_{d,s}\to \phi K_s$}

\begin{figure}[hbt]
\vfill
\begin{center}

\includegraphics[width=0.7\textwidth]{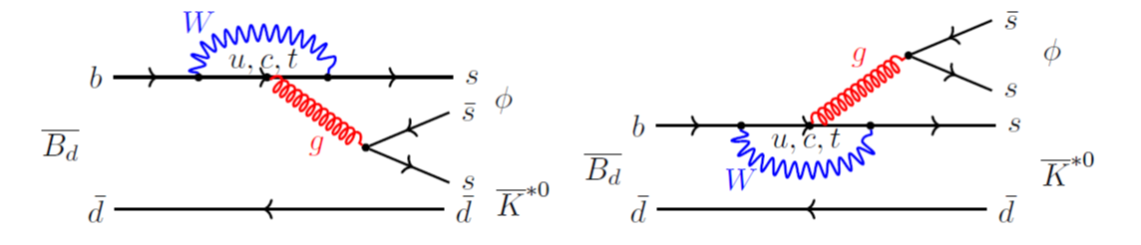}
\includegraphics[width=0.7\textwidth]{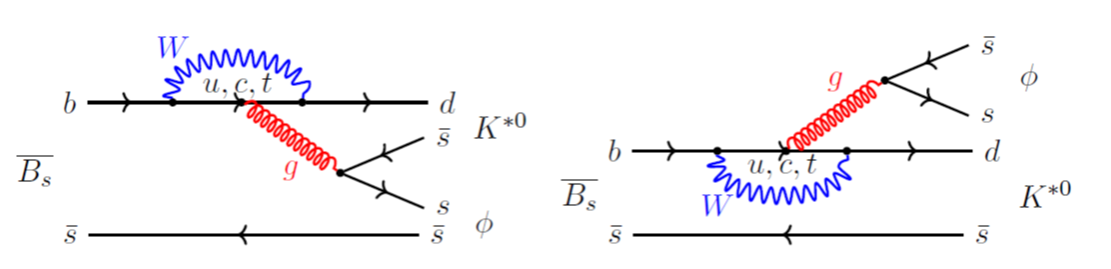}

\caption{\footnotesize \label{fig:B-phiK0} $\overline{B}_{d,s} \to \phi  K^{*0} ( \overline{K}^{*0}) $ decay diagrams. The diagrams for $B_{d,s}\to \phi K_s$ are identical. Not shown here, there are also annihilation diagrams, which are however subleading.}
\end{center}
\vfill

\end{figure}

These decays are pure penguin decays, i.e. no tree diagrams, either color allowed or suppressed, are possible. Figure~\ref{fig:B-phiK0} show the main diagrams.

\noindent Let us concentrate on the case where the $K^{*0}( \overline{K}^{*0})$ decays to the CP eigenstates $K_s \pi^0$. Then both $B_{d,s}$ and $\overline{B}_{d,s}$ can decay to the final state and therefore CP violation occurs through $B-\overline{B}$ mixing.

\be
\begin{array}{ccccl} 
|B_{L(H)}& = & p|B_{d,s}> + (-) q|\overline{B}_{d,s} > \\
\end{array}
\label{eq:Bs_LH}
\ee

\begin{figure}[hbt]
\vfill
\begin{center}

\includegraphics[width=0.75\textwidth]{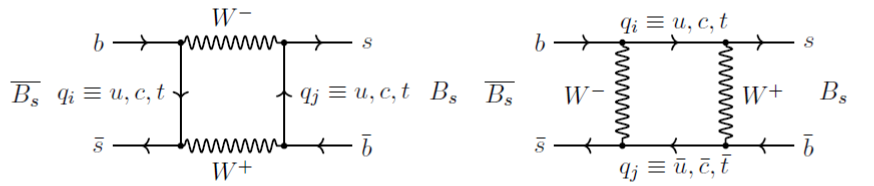}

\caption{\footnotesize \label{fig:Bs_box} The box Feyman diagrams for the $\overline{B_s}-B_s$ mixing. The mixing is dominated by $t-$quark exchange. Similar diagrams are involved for $\overline{B_d}-B_d$ mixing, in which the $s-$quark is replaced by a $d-$quark.}
\end{center}
\vfill

\end{figure}

\noindent In the Standard Model, the box diagrams in Figure~\ref{fig:Bs_box}, which are responsible for $B_s - \overline{B_s} $ mixing, are overwhelmingly dominated by the $t-$quark exchange. 
Thus one can safely use the approximation $\left| q / p \right|_{B_s} \simeq 1$, where $q/p$ is given by a ratio of CKM elements $V_{tb}^* V_{ts}$. This approximation is good at the sub per mille level. Similar diagrams are involved for  $B_d - \overline{B_d} $ mixing and thus one obtains  $\left| q / p \right|_{B_d} \simeq 1$, where $q/p$ is given by a ratio of CKM elements $V_{tb}^* V_{td}$.

\be
\begin{array}{ccccl} 
\left({q \over p} \right)_{B_s} & = &  -\left(\sqrt{ {M^*_{12} \over M_{12} } }\right)_{B_s} & \simeq &  -{V_{tb}^* V_{ts} \over V_{tb} V_{ts}^*} \\
\end{array}
\label{eq:Bs_mixing_s}
\ee
and 
\be
\begin{array}{ccccl} 
\left({q \over p} \right)_{B_d} & = &  -\left(\sqrt{ {M^*_{12} \over M_{12} } }\right)_{B_d} & \simeq &  -{V_{tb}^* V_{td} \over V_{tb} V_{td}^*} \\
\end{array}
\label{eq:Bs_mixing_d}
\ee

  \be
\begin{array}{ccccccc} 
\lambda_{d,s} (f) &=& \left({q \over p} \right)_{B_{d,s} }{ < f |\overline{B}_{d,s} > \over < f |B_{d,s} > } & , &  \overline{\lambda}_{d,s} (f) &=& \left({p \over q} \right)_{B_{d,s} } { < f |{B}_{d,s} > \over < f |\overline{B}_{d,s} > } \\
\end{array}
\label{eq:Bs_lambda_1}
\ee
writing $\rho_{d,s} =\left|\lambda_{d,s}(f) \right|$ and assuming  top dominance, one has 
\be
\begin{array}{ccccccc} 
\lambda_{d,s} (f) &=& \rho_{d,s} e^{i(\phi_{CKM})} & , &  \overline{\lambda }_{d,s}(f) &=&\frac {1}{\rho_{d,s}} e^{-i(\phi_{CKM})}\\
\end{array}
\ee
\vskip 10pt

\noindent where $\phi_{CKM}$ is the CKM phase.

\noindent The time dependent distributions for these decays read :
\be
\begin{array}{ccccl} 
\Gamma(\overline{B}_q(t)\to f) = N_{q,f}  |A_{q,f}|^2 \left[\frac{1+\rho_q^2}{2}\right] e^{-\Gamma_q t} \times \\ 
\left[ \cosh\frac{\Delta \Gamma_q t}{2} - A_{CP}^{dir}\cos(\Delta m_q t) + A_{\Delta\Gamma_q}\sinh\frac{\Delta \Gamma_q t}{2} - A_{CP}^{mix}\sin(\Delta m_q t)\right] \\
\Gamma({B}_q(t)\to f) = N_{q,f}  |A_{q,f}|^2 \left[\frac{1+\rho_q^2}{2}\right] e^{-\Gamma_q t} \times \\ 
\left[ \cosh\frac{\Delta \Gamma_q t}{2} + A_{CP}^{dir}\cos(\Delta m_q t) + A_{\Delta\Gamma_q}\sinh\frac{\Delta \Gamma_q t}{2} + A_{CP}^{mix}\sin(\Delta m_q t)\right] \\\end{array}
\label{eq:time}
\ee
\
\noindent with 

\be
\begin{array}{ccccl} 
A_{CP}^{dir} = \frac{1-\rho_q^2}{1+\rho_q^2}& , &  A_{\Delta\Gamma_q} =-\frac{2Re\lambda_{q,f}}{1+\rho_q^2}& , & A_{CP}^{mix} = -\frac{2Im\lambda_{q,f}}{1+\rho_q^2}
\\\end{array}
\label{eq:def}
\ee

\vskip 10pt
\noindent For the decays $\overline{B}_{d,s} \to \phi (K_s\pi^0)_{K^{*0}}$, the product of the CKM elements is invariant. Namely with top dominance, one gets for $B_s$ :

\be
\begin{array}{ccccl} 
- {V_{tb}^* V_{ts} \over V_{tb} V_{ts}^*}\times {V_{tb} V_{td}^* \over V_{tb}^* V_{td}}\times {V_{us}^* V_{ud} \over V_{us} V_{ud}^*}  & =
& -{V_{ud} V_{us}^* \over V_{td} V_{ts}^*} \times {V_{td}^* V_{ts} \over V_{ud}^* V_{us}} \\ 
=
\left|{V_{ud} V_{us}^* \over V_{td} V_{ts}^*}\right| \times  \left|{V_{td}^* V_{ts} \over V_{ud}^* V_{us}}\right| e^{i\phi_{CKM}} & \\
\end{array}
\label{eq:CKM_Bs_0}
\ee
\noindent with
\be
 \phi_{CKM} = \pi +2\alpha_{ds}
\label{eq:CKM_Bs_01}
\ee

\vskip 10pt

\noindent similarly, one gets for $B_d$ :

\be
\begin{array}{ccccl} 
- {V_{tb}^* V_{td} \over V_{tb} V_{td}^*}\times {V_{tb} V_{ts}^* \over V_{tb}^* V_{ts}}\times {V_{ud}^* V_{us} \over V_{us} V_{ud}^*}  & =
& -{V_{us} V_{ud}^* \over V_{ts} V_{td}^*} \times {V_{ts}^* V_{td} \over V_{us}^* V_{ud}} \\ 
=
\left|{V_{us} V_{ud}^* \over V_{ts} V_{td}^*}\right| \times  \left|{V_{ts}^* V_{td} \over V_{us}^* V_{ud}}\right| e^{i\phi_{CKM}} & \\
\end{array}
\label{eq:CKM_Bd_0}
\ee
\noindent with
\be
 \phi_{CKM} = \pi -2\alpha_{ds}
\label{eq:CKM_Bd_01}
\ee

\vskip 10pt
\noindent Therefore with both $\overline{B}_{s}$ and $\overline{B}_d$ decays to $\phi K^{*0}(\overline{K}^{*0})$, where $K^{*0}(\overline{K}^{*0})$ decays to $K_s\pi^0$, one measures the angle $\alpha_{ds}$ of the 3$^{nd}$ Unitarity Triangle in Fig.~\ref{fig:UT654}. The same result is obtained with the decays $\phi K_s$. With top dominance, the sum of the CKM phases   for $\overline{B}_s\to \phi K_s $ and $\overline{B}_d\to \phi K_s $ is $2\pi$. {\it It is therefore very important to make this measurement with both $B_{s}$ and $B_d$}. This will enable to be sensitive to new physics if the results  show different values for $\alpha_{ds}$.

\noindent Note also that if unitarity with 3 families holds, as in the SM, one has 
\be
\begin{array}{ccccl} 
\alpha_{ds}  & = & \beta \ +\ \beta_s \ - \gamma_{ds}
\end{array}
\label{eq:CKM_alpha_s}
\ee
\noindent  It is thus an indirect measurement of $\beta_s$ since $\beta - \gamma_{ds}$ is know from the process $\overline{B}_d\to J/\psi K_s$. This indirect measurement is not competitive with the direct measurement using $\overline{B}_s\to J/\psi \phi$, however it allows to check the consistency of the SM.

\subsection{Expectation with QCD Factorization}
Table \ref{tab:B_decays} shows the experimental data for the modes  $\overline{B}_{d}\to \phi \overline{K}^{(*)0},\overline{B}_{s}\to \phi{K}^{(*)0}$.

\begin{table}[htb]
\centering
\footnotesize
$$ \begin{tabular}{lcccccc}

$\displaystyle {\mathrm {{B}\ decay}} $ &
$\displaystyle {\mathrm {Br} (\times10^{-6})} $ & 
$\displaystyle {f_L} $ &
$\displaystyle {f_\parallel} $  &
$\displaystyle {f_\perp} $ \\
\hline \hline

$\displaystyle \overline{B}^0\to \overline{K}^{0}\phi$ &
$\displaystyle {\mathrm {7.3\pm0.7}}$ &
$\displaystyle {\mathrm {n/a}} $ &
$\displaystyle {\mathrm {n/a}} $ &
$\displaystyle {\mathrm {n/a}} $ \\

$\displaystyle \overline{B}^0\to \overline{K}^{*0}\phi$ &
$\displaystyle {\mathrm {10.0\pm0.5}}$ &
$\displaystyle {\mathrm {0.497 \pm 0.017}} $ &
$\displaystyle {\mathrm {{\it 0.279\pm 0.023}}} $ &
$\displaystyle {\mathrm {0.224\pm 0.015}} $ \\

$\displaystyle \overline{B}_s\to \phi{K}^{0}$ &
$\displaystyle {\mathrm {1.3\pm0.6}^*}$ &
$\displaystyle {\mathrm {n/a}} $ &
$\displaystyle {\mathrm {n/a}} $ &
$\displaystyle {\mathrm {n/a}} $ \\

$\displaystyle \overline{B}_s\to \phi{K}^{*0}$ &
$\displaystyle {\mathrm {1.14\pm0.30}}$ &
$\displaystyle {\mathrm {0.51 \pm 0.17}} $ &
$\displaystyle {\mathrm {0.21\pm 0.11}} $ &
$\displaystyle {\mathrm {{\it 0.28\pm 0.20 }}} $ \\

\hline\hline

\hline

\end{tabular}   $$

\label{tab:B_decays}
\caption{\footnotesize \label{tab:B_decays} Branching fractions, $f_L$, $f_\parallel$ and $f_\perp$ from the PDG~\cite{pdg:1}. $f_L$, $f_\parallel$ and $f_\perp$ are the longitudinal, parallel and perpendicular polarization fractions, respectively. Statistical and systematic errors have been added in quadrature. The values in italic are not measured directly but are deduced from $f_L+ f_\parallel +f_\perp = 1$. $^*$This Branching fraction includes all ${\rm K^+K^- K^0}$ decays.  }
\end{table}

\noindent In $\rm \overline{B}\to V_1V_2$ decays, one is dealing with 3 helicity amplitudes.
\begin{equation}
\begin{array}{lclcccc} 
{\cal \overline{A}}_0 & = & A[\overline{B}\to V_1(0)V_2(0)] & , & {\cal \overline{A}}_\pm & = &   A[\overline{B}\to V_1(\pm)V_2(\pm)]    \\
\end{array}
\label{eq:A}
\end{equation}

\vskip 10pt
\noindent Moving from the helicity representation to the transversity one, one gets :
\begin{equation}
\begin{array}{lclcccc} 
{\cal  \overline{A}}_L& \equiv & {\cal \overline{A}}_0 \\
\\
{\cal \overline{A}}_\parallel& = & \frac{{\cal \overline{A}}_+ +  {\cal \overline{A}}_-}{\sqrt{2}} & , & {\cal \overline{A}}_\perp& = & \frac{{\cal \overline{A}}_+ -  {\cal \overline{A}}_-}{\sqrt{2}}  \\
\end{array}
\label{eq:A_trans}
\end{equation}

\noindent with the corresponding transversity rate fractions $f_L$, $f_\parallel$ and $f_\perp$ satisfying
\begin{equation}
\begin{array}{lclcccc} 
f_L+f_\parallel +f_\perp = 1  \\
\end{array}
\label{eq:f_eq}
\end{equation}

\noindent At $q^2=0$ and in the heavy quark limit and large recoil energy for the light meson,
\vskip 10pt 
\begin{equation}
\begin{array}{cccccc} 
{\cal \overline{A}}_+\simeq 0
\end{array}
\label{eq:A=0}
\end{equation}
\noindent and one then has 
\begin{equation}
\begin{array}{cccc} 
{\cal \overline{A}}_\parallel \simeq   -{\cal \overline{A}}_\perp& \simeq & \frac{{\cal \overline{A}}_-}{\sqrt{2}}  \\
\\
|{\cal \overline{A}}_\parallel|^2 + |{\cal \overline{A}}_\perp|^2& \simeq & |{\cal \overline{A}}_-|^2 \equiv |{\cal \overline{A}}_T|^2\\
\end{array}
\label{eq:F}
\end{equation}
\vskip 10pt
\noindent where the subindex $T$ stands for transverse. Finally, in the SM ($V-A$), one gets
\begin{equation}
\begin{array}{cccc} 
f_\parallel & \simeq&  f_\perp
\end{array}
\label{eq:F=F}
\end{equation}
\vskip 10pt
\noindent As can be seen in Table~\ref{tab:B_decays}, equation~(\ref{eq:F=F}) seems to be verified within experimental errors.
\noindent We examine now whether $\phi_{L,\parallel, \perp}^{CKM}$ differ significantly when $c,u$ quarks are considered in the loops using QCD Factorization~\cite{AO:1}. 
As mentioned above, equations~(\ref{eq:CKM_Bs_0}) and~(\ref{eq:CKM_Bd_0}) hold exactly when top dominance is assumed. Table~\ref{tab:QCDF} shows the expected values for $\lambda_{L,\parallel, \perp}$ and $\phi_{L,\parallel, \perp}^{CKM}$ using QCD Factorization.
As it can be observed, the measurement of $2\alpha_{d,s}$ still holds within the theoretical errors. However one notes also that $|\lambda_{L,\parallel, \perp}|$ is different from 1 for the decay $\overline{B}_s\to \phi K^{(*)0}$, due to the presence of direct CP violation effects. This is due to the fact that the CKM elements involved in the leading penguin diagram for the $\overline{B}_s$ decay are of the same order, $\lambda^3$, in contrast to the $B_d$, for which the dominant terms (with $t$ and $c$ exchange) are of order $\lambda^2$ while the other term (with $u$ exchange) is of order $\lambda^4$.

\begin{table}[htb]
\centering
\tiny
$$ \hspace{-1.5cm}
\begin{tabular}{cccccc}
\hline

$\displaystyle {\mathrm {Decay}} $ &
$\displaystyle \overline{B}_d\to \phi \overline{K}^0$ &
$\displaystyle  \overline{B}_d\to \phi \overline{K}^{*0}$ &
$\displaystyle \overline{B}_s\to \phi K^0$ &
$\displaystyle  \overline{B}_s\to \phi K^{*0}$ \\

\hline \hline

$\displaystyle {\mathrm {|\lambda_L|}} $ & 
$\displaystyle 1.017 \pm 0.005$ &
$\displaystyle 1.016\pm0.005 $ &
$\displaystyle 0.743\pm 0.075 $ &
$\displaystyle 0.746\pm 0.069 $ &
\\ 
$\displaystyle {\mathrm {\phi_L^{CKM}}} $ &
$\displaystyle {\mathrm {\pi - 2\alpha_{ds}+0.004\pm 0.005}} $ &

$\displaystyle {\mathrm {\pi - 2\alpha_{ds}+0.003\pm 0.006}} $ &
$\displaystyle {\mathrm {\pi + 2\alpha_{ds}-0.030\pm 0.121}} $  & 
$\displaystyle {\mathrm {\pi + 2\alpha_{ds}-0.021\pm 0.105}} $ 
\\ 

$\displaystyle {\mathrm {|\lambda_\parallel|}} $ &
$\displaystyle {\mathrm {-}} $ &
$\displaystyle 1.005\pm 0.001  $ & 
$\displaystyle - $ & 
$\displaystyle {\mathrm {0.917\pm 0.021}} $ 
\\ 
$\displaystyle {\mathrm {\phi_\parallel^{CKM}}} $ &
$\displaystyle {\mathrm {-}} $ &
$\displaystyle {\mathrm {\pi - 2\alpha_{ds}-0.002\pm 0.002}} $ &
$\displaystyle {\mathrm {-}} $  & 
$\displaystyle {\mathrm {\pi + 2\alpha_{ds}-0.028\pm 0.035}} $ \\

$\displaystyle {\mathrm {|\lambda_\perp|}} $ &
$\displaystyle {\mathrm {-}} $ &
$\displaystyle 1.005\pm 0.001  $ & 
$\displaystyle - $ & 
$\displaystyle {\mathrm {0.917\pm 0.021}} $ \\

$\displaystyle {\mathrm {\phi_\perp^{CKM}}} $ &
$\displaystyle {\mathrm {-}} $ &
$\displaystyle {\mathrm {\pi -2\alpha_{ds}-0.002\pm 0.002}} $ &
$\displaystyle {\mathrm {-}} $  & 
$\displaystyle {\mathrm {\pi +2\alpha_{ds}-0.028\pm 0.035}} $ 

\\ \hline

\end{tabular}   $$

\label{tab:QCDF}
\caption{\footnotesize \label{tab:QCDF} The expected values of $|\lambda_{L,\parallel , \perp}|$ and $\phi_{L,,\parallel , \perp}^{CKM}$ for $\overline{B_d}$ and $\overline{B_s}$ decays wihin QCD factorization.}
\end{table}

\section{Detector simulation}
\subsection{Generic detector resolutions}
We consider a typical FCC-ee detector in order to study the acceptance efficiency as well as  the momentum, mass and vertex resolutions for the charged tracks. More precisely, a complete tracking simulation including multiple scattering is carried out for a large set of momenta and polar angles to determine the momentum resolution and angular resolutions. For a fast simulation, we then parametrize the resolutions using this set of data. The energy and angular parametrization for photons and electrons assumes a crystal type electromagnetic calorimeter and we use typical conservative resolutions. In summary the detector resolutions are listed in~(\ref{eq:track_resolution}),

\be
\footnotesize
\begin{array}{lccl} 
\mathrm{Acceptance :}& |\cos \theta|&<&0.95\\
\hline
\mathrm{Charged \ particles :} & \\
\mathrm{ p_T\ resolution :}& {\sigma (p_T) \over p_T^2}  & = & 2. \times 10^{-5} \ \oplus \ {1.2 \times 10^{-3}\over p_T \sin \theta}\\
\mathrm{ \phi , \theta \ resolution :}& \mathrm{\sigma (\phi , \theta) \ \mu rad } & = &  18 \ \oplus \ {1.5 \times 10^{3} \over p_T\sqrt[3]{\sin \theta} }\\
\mathrm{Vertex \ resolution :}& \mathrm{\sigma (d_{Im}) \ \mu m} & = &  1.8 \ \oplus \ {5.4 \times 10^{1} \over p_T\sqrt{\sin \theta} }\\
\hline
\mathrm{e,\gamma \ particles :} & \\
\mathrm{Energy\ resolution:}&{\sigma (E) \over E}  & = & {5 \times 10^{-2} \over \sqrt{E}} \ \oplus \ 5 \times 10^{-3}\\
\mathrm{EM \ \phi , \theta \ resolution :}& \mathrm{\sigma (\phi , \theta) \ m rad } & = &  {7 \over \sqrt{E} }\\
\hline \hline
\end{array}
\label{eq:track_resolution}
\ee

\noindent where $\theta, \phi $ are the particles' polar and azymutal angles respectively, $p_T$ (in GeV) the track transverse momentum, $E$ the $e^\pm\ ,\gamma$ energy and $\mathrm{d_{Im}}$ the tracks' impact parameter.
In addition to this parameterised detector response, we have also used Monte-Carlo events processed through a DELPHES~\cite{de_Favereau_2014} simulation of the IDEA detector concept~\cite{fccee:3}; more details will be given in Section 5.2.

\subsection{Vertex resolution at FCC-ee}

\begin{figure}[hbt]
\vfill
\begin{center}

\includegraphics[width=0.60\textwidth]{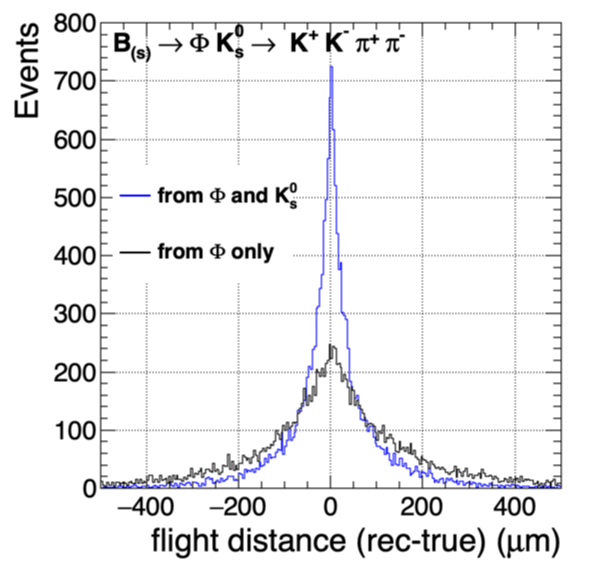}

\caption{\footnotesize \label{fig:phi-resol} The resolution on the $\overline{B}_s$ flight distance using the vertex $\phi -K_s$}
\end{center}
\vfill

\end{figure}

\begin{table}[htb]
\centering
\footnotesize
$$ \begin{tabular}{cccc}
\hline
&  $\displaystyle {\mathrm {\int L = 150\ ab^{-1}}}$   &  \\

$\displaystyle {\mathrm {\sigma (e^+e^- \to Z )}} $ &
$\displaystyle {\mathrm {number}} $ &
$\displaystyle {\mathrm {f(Z\to \overline{B_d})}} $ &
$\displaystyle {\mathrm {f(Z\to \overline{B_s})}} $ \\

$\displaystyle {\mathrm {nb}} $ & 
$\displaystyle {\mathrm {of \ Z} } $ &
$\displaystyle {\mathrm{}} $ &
$\displaystyle {\mathrm{ }} $\\ 
\hline \hline 

$\displaystyle \sim 42.9$ &
$\displaystyle {\mathrm {\sim 6.4\ 10^{12}}}$ &
$\displaystyle {\mathrm {0.06}} $ &
$\displaystyle 0.0159 $\\ 
\hline
& & & \\

$\displaystyle {\mathrm {Decay}} $ &
$\displaystyle {\mathrm {Final}} $ & 
$\displaystyle {\mathrm {Number \ of}} $ &
$\displaystyle {\mathrm {Number \ of}} $ \\

$\displaystyle {\mathrm {Mode}}  $ &
$\displaystyle {\mathrm {State} } $ &
$\displaystyle {\mathrm{\overline{B_d} \ decays}} $ &
$\displaystyle {\mathrm{\overline{B_s} \ decays}} $ \\ 
\hline \hline
& $\displaystyle {\mathrm {CP\ eigenstates}}$   &  \\

$\displaystyle \phi K^0$ &
$\displaystyle {\mathrm {K^+K^-(\pi^+\pi^-)_{K_s}}}$ &
$\displaystyle {\mathrm {\sim 4.9\ 10^5}} $ &
$\displaystyle \sim 1.7\ 10^4 $\\ 
$\displaystyle  \phi K^{*0}$ &
$\displaystyle {\mathrm {K^+K^-(\pi^+\pi^-)_{K_s}\pi^0}}$ &
$\displaystyle {\mathrm {\sim 2.1\ 10^5}}$ &
$\displaystyle \sim 6.4 \ 10^3 $\\ 
\hline

\end{tabular}   $$

\label{tab:stat}
\caption{\footnotesize \label{tab:stat} The expected number of produced $\overline{B_d}$ and $\overline{B_s}$ decays to specific modes at FCC-ee at a center of mass energy of 91 GeV over 5 years with 2 detectors. These numbers have to be multiplied by 2 when including $B_d$ and $B_s$ decays. The branching fractions of the PDG~\cite{pdg:1} have been used whenever available, else the expectation from QCD factorization is used.}
\end{table}
\vskip 10pt
The expected number of produced events at FCC-ee are listed in Table~\ref{tab:stat}. In order to study CP violation one needs to carry out a time dependent measurement. It is therefore important to have a good vertex resolution, which is particularly critical for the $B_s$ decays since the oscillation frequency is high. The $B_s$ flight distance resolution at the Z-pole with a FCC detector has been studied in detail for the decay $B_s\to D_s^\pm K^\mp$ in earlier work~\cite{AOP:1} and one finds $\sim 20\mu m$. However in the decay $\overline{B}_s \to \phi K^{(*)0}$, the $K^0 (K^{*0})$ has to decay to $K_s(K_s\pi^0)$ and therefore the resolution on the $\overline{B}_s$ decay vertex is determined by the $\phi K_s$ vertex, which is less precise than  the vertex $D_s^\pm K^\mp$ with $D_s^\pm \to \phi\pi^\pm$ . We have therefore studied the resolution on the $\overline{B}_s$ flight distance using the $\phi$ and the $K_s$. This study uses signal Monte-Carlo events with a $\overline{B_s} \rightarrow \phi K_{s}$ decay processed through DELPHES, and a vertexing software~\cite{BedeschiCode, Franco_neutrals} that handles both charged and neutral particles. An average resolution of $~70\mu m$ is found, see Figure~\ref{fig:phi-resol}. This resolution is, as expected, worse that the $20\mu m$ found with the full vertex information in the $B_s\to D_s^\pm K^\mp$ decays. This is due to the fact that the 2 Kaons from the $\phi$-meson are produced with a small momentum (127 MeV/c) in the $\phi$ center of mass and they are emitted close to each other. The addition of the $K_s$ improves significantly the resolution as it would be $168\mu m$ should one use only the $ \phi$. We remind nevertheless that the average $B_s$ flight distance is $3mm$, hence the resolution of $70\mu m$ is excellent and does not significantly dilute the oscillations.

\section{Background studies}

\noindent The final state $\phi K^0$ includes 2 charged kaons and a $K_s$ while the final state $\phi K^{*0}$ includes in addition a $\pi^0$. Thanks to the excellent PID and the excellent momentum resolution, which are foreseen at the FCC-ee detectors, it is expected that these modes are essentially background free.  We have nevertheless verified this using some exclusive final states processed with the parameterised detector response (Section 5.1), as well as generic $Z \rightarrow b \bar{b}, c \bar{c}$ events generated with PYTHIA and simulated with DELPHES (Section 5.2).

\subsection{Exclusive final states}

\begin{figure}[hbt]
\vfill
\begin{center}

\includegraphics[width=0.49\textwidth]{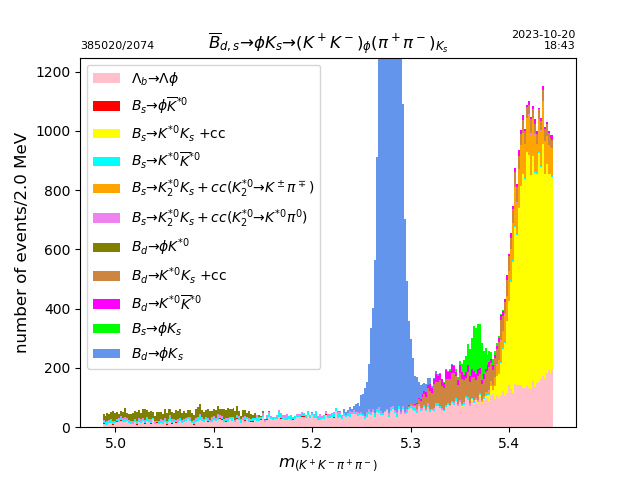}
\includegraphics[width=0.49\textwidth]{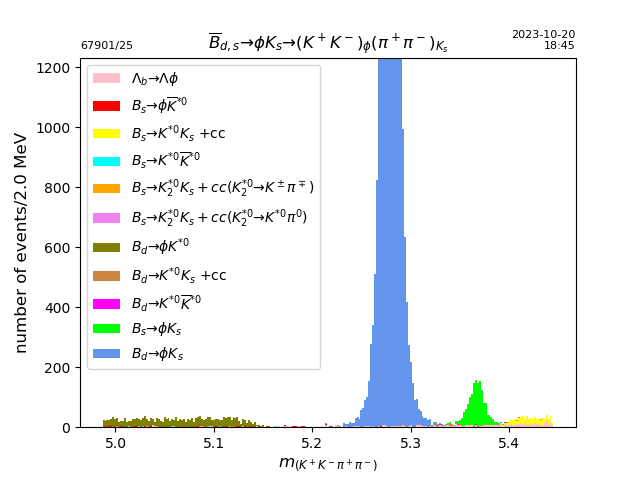}

\caption{\footnotesize \label{fig:bgd1} Reconstructed final states $\overline{B}_{d,s}\to \phi K_s$ for an integrated luminosity of 15 $\rm ab^{-1}$ at the Z-pole. In the left plot,  no cut on the $\rm K^+K^-$ and $\rm \pi^+\pi^-$ mass is applied. The right plot includes the cuts $\rm 1.00\ GeV<M_{K^+K^-}<1.04\ GeV$ and $\rm 0.489\ GeV<M_{\pi^+\pi^-}<0.506\ GeV$. }
\end{center}
\vfill

\end{figure}

There are 3 main categories of exclusive final states that could potential contribute to the background :
\begin{enumerate}
\item Final states with light mesons without long live particles (e.g. $K_s$ or $\Lambda$) such as $\phi \rho^0 \to ( K^+K^-)_{\phi}(\pi^+\pi^-)_{ \rho^0}$ or $\phi f_0(980) \to ( K^+K^-)_{\phi}(\pi^+\pi^-)_{ f_0(980)}$
\item Final states with long live particles (e.g. $K_s$ or $\Lambda$) such as $K^{*0} \overline{K}^0$ +cc or $\Lambda\phi$
\item Final states with $D_s^\pm$ particles decaying to $\phi\pi^\pm$  such as $D_s^\pm \pi^\mp$

\end{enumerate}
\noindent The background in the first category is abundant but  can be easily rejected by requiring the $\pi^+\pi^-$ mass to be around the $K^0$ mass and its vertex to be detached from the $(K^+K^-)_\phi$ vertex. This is discussed above in the vertexing section. This requirement rejects essentially all of this background, it also rejects part of the combinatoric background.
\vskip 10pt
\noindent In the second category, the exclusive final states considered are $B_{d,s}\to K^{*0}K_s$ +cc, $B_{s}\to K^{*0}_2(1430)K_s\to K^{*0}\pi^0K_s$+cc, and $B_{d,s}\to K^{*0}\overline{K}^{*0}$ in which one of the $K^{*0}$ or  $\overline{K}^{*0}$ decays to $K_s\pi^0$ while the other decays to $K^\pm \pi^\mp$. We have also included the decay $\Lambda_b \to \Lambda \phi$. These modes include a charged $\pi$ or a proton  and therefore are not a background, should one have a good Particle Identification (PID) system. Nevertheless, let us assume for now that one does not use PID. We show in Figure~\ref{fig:bgd1} the reconstructed mass of the final state $\overline{B}_{d,s}\to \phi K_s$, i.e. with a wrong assignment of the pion (proton) as a kaon (pion).
 
\vskip 10pt
\noindent The mass resolution of the $\phi K_s$ system is better than 9 MeV. It can be seen that as soon as the mass constraint is used for the $\phi$ and the $K_s$, essentially all background disappears and we are left with clear peaks for $\overline{B}_{d,s}$. Needless to say, should one have a PID system, these exclusive backgrounds would disappear as well, even without cutting on the $\phi$ or $K_s$ mass.

\vskip 10pt
\noindent Finally let us consider the third category, which is potentially dangerous since the final set of particles, $\phi \pi^+\pi^-$, is identical to the one in $\phi K_s$. Indeed the expected rate for $B_s \to D_s^\pm\pi^\mp\to \phi \pi^\pm\pi^\mp$ is about 400 times larger than $B_s\to\phi K_s\to \phi \pi^\pm\pi^\mp$. There are 3 means to reject this background:
\begin{enumerate}
\item Requiring the $\pi^+\pi^-$ mass to be around the $K^0$ mass, 
\item Eliminating events in which the combination of $\phi\pi^\pm$ is around the $ D_s^\pm$ mass,
\item Requiring the $\pi^+\pi^-$ pair to form a good vertex detached from the $\phi$ vertex.

\end{enumerate}

\begin{figure}[hbt]
\vfill
\begin{center}

\includegraphics[width=0.49\textwidth]{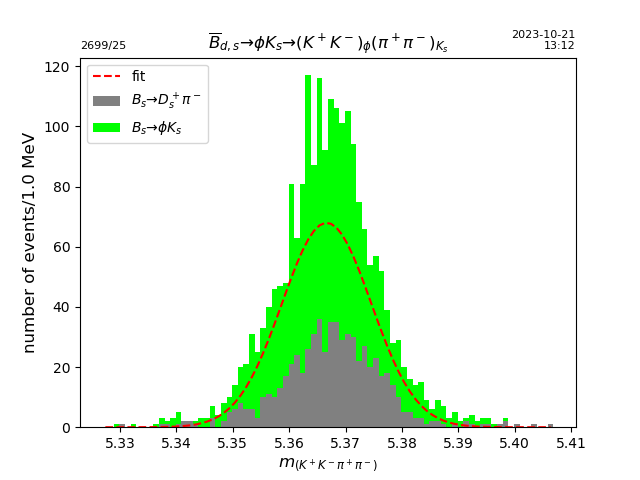}
\includegraphics[width=0.49\textwidth]{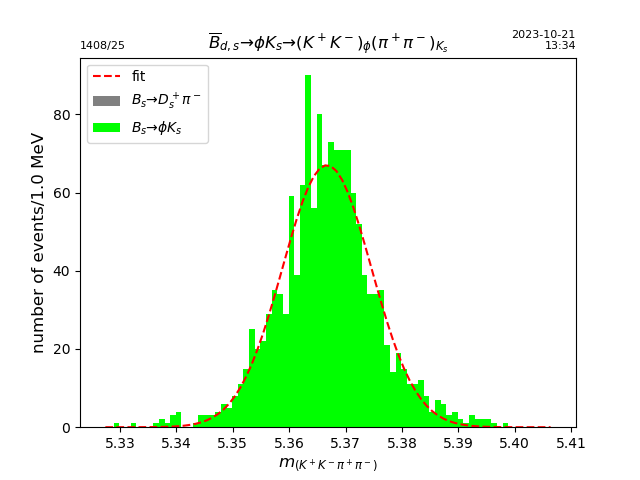}

\caption{\footnotesize \label{fig:bgd2} Reconstructed final states $\overline{B}_{s}\to \phi K_s$ for an integrated luminosity of 15 $\rm ab^{-1}$ at the Z-pole. In the left plot,  cuts on the $\rm K^+K^-$ and $\rm \pi^+\pi^-$ mass are applied but no cut on the $\rm K^+K^-\pi^+$ and  $\rm K^+K^-\pi^-$ combinations. The right plot includes in addition the requirement that no combination  $\rm M_{K^+K^-\pi^\pm}$ is in the range $\rm 1.938 - 1.998\ GeV$. }
\end{center}
\vfill

\end{figure}

\noindent We show the effects of cut 1 and cut 2 in Figure~\ref{fig:bgd2}. The backround $D_s \pi$ is completely eliminated with essentially no event loss for the signal.

\subsection{Generic $b\overline{b},\ c\overline{c}$ events}

Therefore, the main source of background is expected to be of combinatorial origin. Inclusive Monte-Carlo samples of Z $\rightarrow b \bar{b}$ and Z $\rightarrow c \bar{c}$ events have been used to confirm this expectation, and to quantify the level of the combinatoric background. They consist of one billion of $b \bar{b}$ events, and of 500 millions of $c \bar{c}$ events, produced with the PYTHIA~8.306 Monte-Carlo generator~\cite{Bierlich:2022pfr}. Signal events are removed from the inclusive $b \bar{b}$ background sample. They are generated separately, using PYTHIA to simulate the production of a $b \bar{b}$ pair in which one $b$ quark hadronises into a $B_d$ or $B_s$ that decays into $\phi K_s$, while the other $b$-leg fragments and decays inclusively. The $B_{d,s} \rightarrow \phi K_s$ decay chain was performed with the EvtGen~\cite{LANGE2001152} program. 
The generated events were passed through a fast simulation of the IDEA detector~\cite{fccee:3}, which provides resolutions similar to the ones given in Section~4.1. The simulation is based on DELPHES~\cite{de_Favereau_2014}.
In particular, the simulation software that turns charged particles into simulated tracks relies on a full description of the geometry of the IDEA vertex detector and drift chamber. The software accounts for the finite detector resolution and for the multiple scattering in each tracker layer and determines  the (non diagonal) covariance matrix of the helix parameters that describe the trajectory of each charged particle. This matrix is then used to produce a smeared 5-parameters track, for each charged particle emitted within the angular acceptance of the tracker. Finally, the events were subsequently analysed within the FCCAnalyses framework~\cite{FCCAnalyses}.  \\

\noindent The reconstruction of signal candidates starts with the identification of the ``primary tracks'', that can be fit to a primary vertex\footnote{A simple iterative algorithm is used here. In a first step, all tracks are fit to a common vertex, using a constraint given by the beam-spot size. The track that gives the largest contribution to the $\chi^2$ of the fit is removed, and the remaining tracks are fit again. The procedure is repeated until the $\chi^2$ contribution of each track is below a given cut.}, and, consequently, of the ``secondary tracks''. Moreover, all reconstructed particles are used to determine the thrust axis, and the plane orthogonal to this axis and containing the interaction point divides each event in two hemispheres. \\
Pairs of opposite-charge secondary tracks that belong to a same hemisphere are fit to a common vertex. Pairs for which the vertex fit has a good $\chi^2$ (a rather loose cut, $\chi^2 < 10$, being used here), and whose invariant mass (determined from the tracks' momenta at the fitted vertex) is within $1.00$ and $1.04$ GeV ($448$ and $548$ MeV) define $\phi$ ($K_s$) candidates. This set of cuts appears with the label ``$1$'' in Tab.~\ref{tab:full_selection}, which summarises all selection criteria. Only $K_s$ candidates that decay within $1.5$~m from the interaction point are selected for further analysis (cut $2$); this cut removes $K_s$ candidates made of short tracks, prone to large measurement uncertainties. For pairs of $\phi$ and $K_s$ candidates that belong to a same hemisphere (cut $3$), a vertex is fit from the two tracks that make the $\phi$ candidate and from the trajectory of the neutral $K_s$. The standalone vertex fit algorithm~\cite{BedeschiCode} used in this analysis is available in the distribution of the DELPHES package, and its recent extension to allow neutral particles to be included in the fit is described in~\cite{Franco_neutrals}. Pairs with an invariant mass between $5.33$ and $5.41$ GeV ($5.24$ and $5.32$ GeV), and for which the normalised $\chi^2$ of this latter vertex fit is smaller than $7.5$, define $B_s$ ($B_d$) candidates (cut $4$). Events containing at least one such candidate are kept for further analysis. \\
\begin{figure}[tbh]
  \centering
  \begin{tabular}{cc}
  \includegraphics[width=0.40\columnwidth]{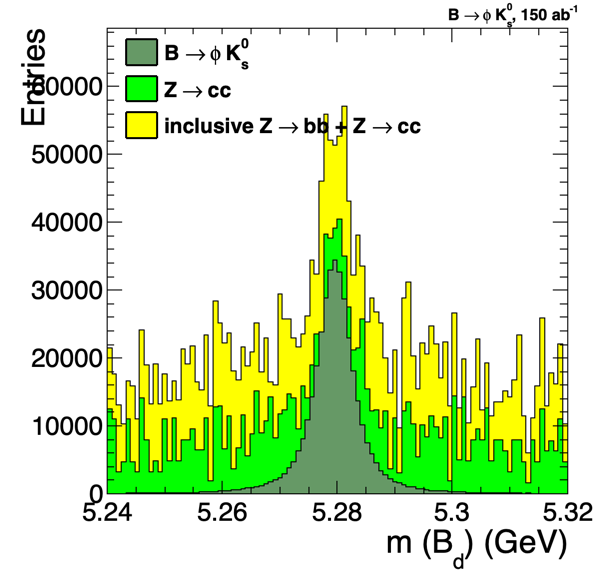}  &
  \includegraphics[width=0.40\columnwidth]{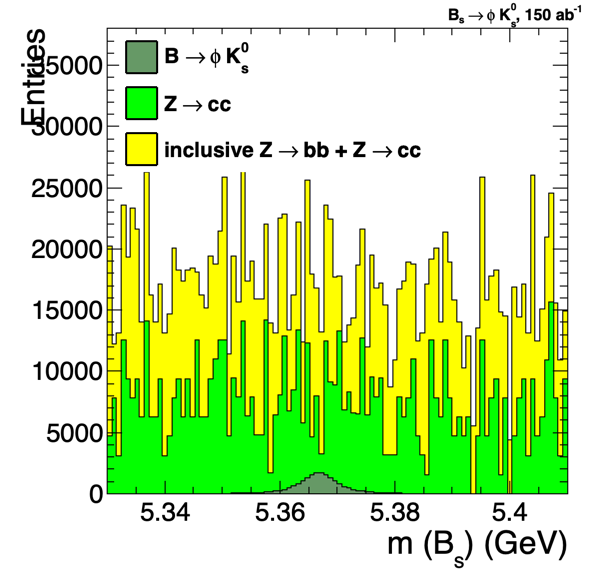} \\
  \includegraphics[width=0.40\columnwidth]{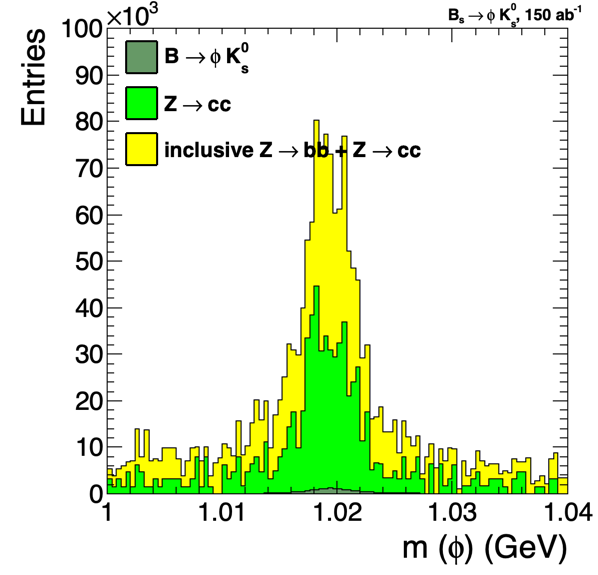} &
  \includegraphics[width=0.40\columnwidth]{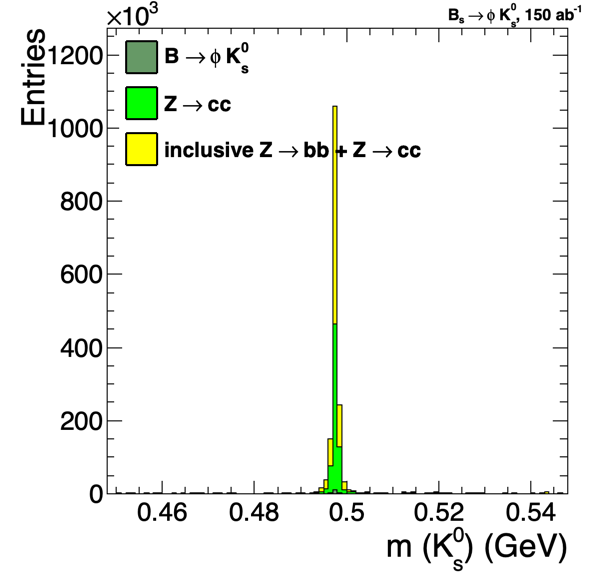}
  \end{tabular}
  \caption{\footnotesize Top: Distribution of the mass of reconstructed $B_d$ (left) and $B_s$ (right) candidates prior to any selection cut. The histograms corresponding to the signal, to the $Z \rightarrow c \bar{c}$ background and to the $Z \rightarrow b \bar{b}$ background are stacked on top of each other. Bottom: Distribution of the mass of the $\phi$ candidate (left) and of the $K_s$ candidate that make the $B_s$ candidate. 
  }
  \label{fig:masses_nocut}
\end{figure}

\begin{figure}[thb]
  \centering
  \begin{tabular}{cc}
  \includegraphics[width=0.40\columnwidth]{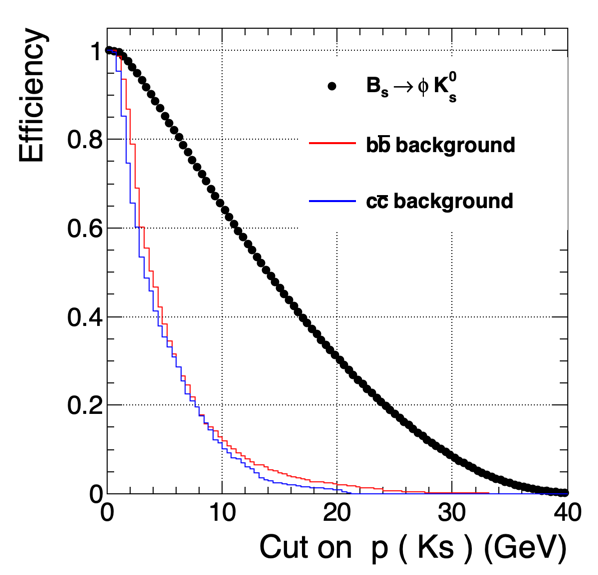}  &
  \includegraphics[width=0.40\columnwidth]{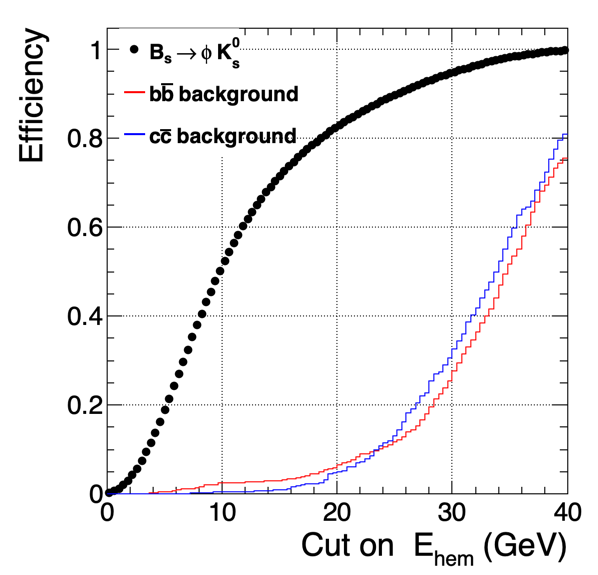} 
  \end{tabular}
  \caption{\footnotesize Efficiency of a lower cut on the $B_s$ momentum (left) and of an upper cut on the energy reconstructed in the signal hemisphere after subtracting that of the $B_s$ candidate (right). The dots show the $B_s$ signal efficiency, and the red and blue lines the efficiency over $b \bar{b}$ and $c \bar{c}$ background events. The efficiencies are determined on top of the basic reconstruction criteria used in Fig.~\ref{fig:masses_nocut}.
  }
  \label{fig:kinemCuts_efficiency}
\end{figure}

\noindent In a first step, a perfect PID is assumed and the Monte-Carlo information is used to demand that the legs of the $\phi$ and $K_s$ candidates that make the $B$ candidate be kaons and pions, respectively (cut $5$). At this stage, about $57 \%$ of signal events are selected, the loss being mainly due to the $K_s$ acceptance. About $2$ in $10^6$ $b \bar{b}$ events contain a $B_{(s)} \rightarrow \phi K_s$ candidate, the rate for $c \bar{c}$ events being similar.  The background is large compared to the signal, in particular for the small $B_s$ signal, as shown in the top plots of Fig~\ref{fig:masses_nocut}. The $\phi$ and $K_s$ candidates that make $B$ candidates are usually genuine $\phi$ and $K_s$ particles, as shown by the lower plots of the same figure. The following cuts are applied to $B_{s}$ candidates in order to suppress the background due to the exclusive processes considered in the previous section:
\begin{itemize}
 \item the distance between the $K_s$ decay vertex and the $B_{s}$ decay vertex is required to be larger than $1$~mm (cut $6$);
 \item three-tracks vertex fits are run, from the two tracks that make the $\phi$ candidate and from each other track that belongs to the same hemisphere as the $\phi$. If there is a track for which the resulting vertex has an acceptable $\chi^2$ and a mass\footnote{When determining the vertex mass, the track that does not come from the $\phi$ is given the pion mass, unless it comes from a muon or an electron, in which case the corresponding lepton mass is used.} below $1.986$~GeV (the nominal $D_s$ mass plus about twice the mass resolution), the $B$ candidate is rejected (cut $7$).
\end{itemize}
The latter cut efficiently removes $D_s \rightarrow \phi \pi$, $D_s \rightarrow \phi \mu \nu_{\mu}$ and $D_s \rightarrow \phi e \nu_{e}$ events, as well as $D_s \rightarrow \phi \pi + X$ events, at the price of a relative efficiency loss of $15 \%$ on the signal.
\begin{figure}[thb]
  \centering
  \begin{tabular}{cc}
  \includegraphics[width=0.40\columnwidth]{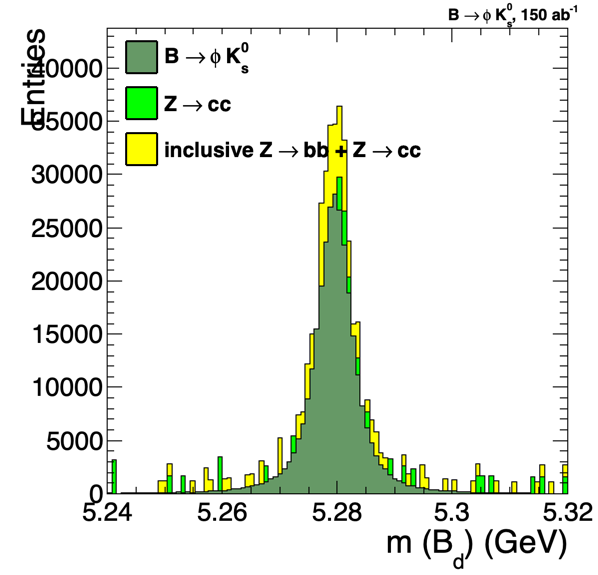}  &
  \includegraphics[width=0.40\columnwidth]{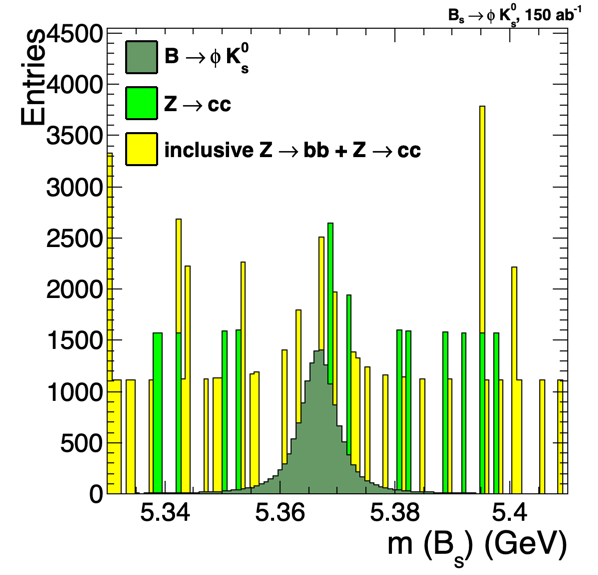} 
  \end{tabular}
  \caption{\footnotesize Distribution of the mass of reconstructed $B_d$ (left) and $B_s$ (right) candidates after the cuts designed against the exclusive processes considered in Section 5.1.  
  }
  \label{fig:masses_against_exclusive}
\end{figure}
The mass distribution of the $B_s$ and $B_d$ candidates passing these cuts are shown in Fig.~\ref{fig:masses_against_exclusive}. For the $B_d$ signal, the purity is now quite good. The background contribution that is seen to peak at the $B$ mass is due to $B \rightarrow f_0(980) K_s$ decays\footnote{With the default PYTHIA settings used here, the mass of the $f_0$ is set to $1$~GeV and its width to $50$~MeV. Both the width, and the branching fraction of the $f_0$ into $K^+ K^-$, are actually very poorly known and, with the FCC data, the contributions from $B \rightarrow \phi K_s$ and from $B \rightarrow f_0 K_s$ will have to be fitted together.}, followed by a decay into $K^+ K^-$ of the $f_0(980)$. This contribution can be limited with a tighter lower cut on the mass of the $\phi$ candidate (cut $(8a)$). On the other hand, the much smaller $B_s$ signal still suffers from a large background. 
It is clearly visible from the $B_s$ mass plot that a much higher Monte-Carlo statistics would be needed in order to properly study the background. Assuming that the mass distribution of the background is flat in the range depicted in the plot, for an integrated luminosity of $150$~ab$^{-1}$, about $11000$  candidates are expected from $b \bar{b}$ background events under the $B_s$ mass peak, with a $16 \%$ uncertainty, and about $5100$ from $c \bar{c}$ events, with a $28\%$ uncertainty. In comparison, $16000$ signal candidates are expected for this luminosity. An investigation of the remaining background to the $B_s$ signal confirms the combinatorial origin of the background at this stage; in the very large majority of the events, either the $\phi$ or the $K_s$ is produced during the fragmentation, while the other meson is coming indeed from the decay of a heavy-flavour hadron. This background can be suppressed by kinematic cuts, since the mesons produced during the fragmentation are usually very soft (cuts $9$). A loose cut on the momentum of the $B_s$ candidate, and on the energy reconstructed in the signal hemisphere after removing that of the $B_s$ candidate, also improves efficiently the signal-to-background ratio, as shown in Fig.~\ref{fig:kinemCuts_efficiency} (cuts $10$). 

\begin{table}[htb]
    \centering
    \tiny
$$   \hspace{-0.cm} \begin{tabular}{|l|cl|}
      \hline
      $K_s$ leg:  & (1) & vertex $\chi^2 < 10$ \\
                  & (1) & $448 < m < 548$~MeV \\
                  & (2, 6) & flight distance $ > 1$~mm and $ < 1.5$~m \\
                  & (9) & $ p > 1.5$~GeV \\ \hline
      $\phi$ leg: & (1) & vertex $\chi^2 < 10$ \\
                  & (1, 8a, 8b) & $ 1.01 < m < 1.03$~GeV \\
                  & (9) & $ p > 1$~GeV \\ \hline 
    \multicolumn{3}{|c|}
    {(5) perfect charged hadron PID for the $K_s$ and $\phi$ tracks}  \\ \hline
      $B_{d,s}$ candidate: & (3) & the four tracks belong to the same hemishpere \\
                           & (4) & vertex $\chi^2 < 7.5$ \\
                           & (4) & $ 5.24 < m < 5.32$~GeV for $B_d$ \\
                           & (4) & $5.33 < m < 5.41$~GeV  for $B_s$ \\
                           & (10) & $ p > 10$~GeV \\ \hline
      Energy in signal hemisphere:  & (10) &   below $28$~GeV \\
    (without the $B$ candidate)     & &      \\ \hline
      \multicolumn{3}{|c|}
      {(7) no reconstructed $D_s^{\pm} \rightarrow \phi + {\rm {track}} + X$ in signal hemisphere.}    \\  \hline
                            
    \end{tabular} $$
    \caption{\footnotesize Summary of the final selection cuts used in this analysis. The labels allow to retrieve easily where the corresponding cuts are introduced in the text.}
    \label{tab:full_selection}
\end{table}
\noindent Another small component to the background comes from non-resonant decays $B_s \rightarrow K^+ K^- K_s$, where the $K^+ K^-$ pair does not come from a $\phi$ meson; it can be further reduced with a tighter cut on the mass of the $\phi$ candidate (cut $8b$).

\begin{figure}[thb]
  \centering
  \begin{tabular}{cc}
  \includegraphics[width=0.40\columnwidth]{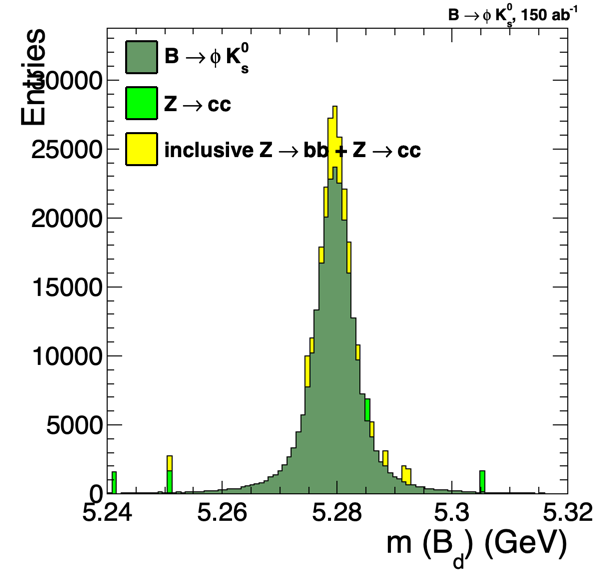}  &
  \includegraphics[width=0.40\columnwidth]{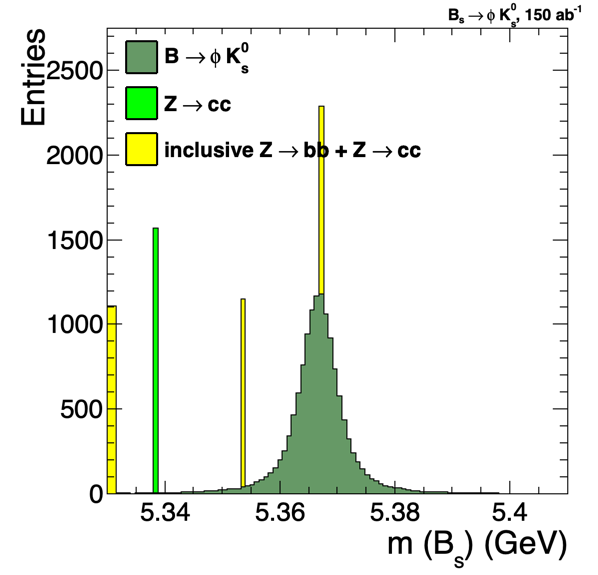} \\
  \includegraphics[width=0.40\columnwidth]{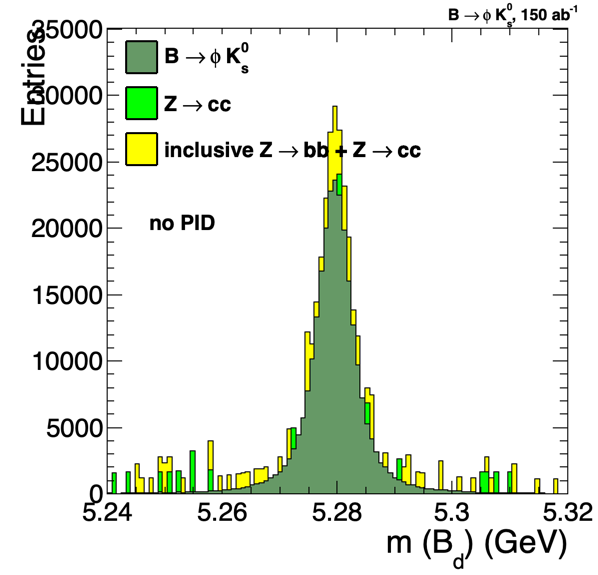}  &
  \includegraphics[width=0.40\columnwidth]{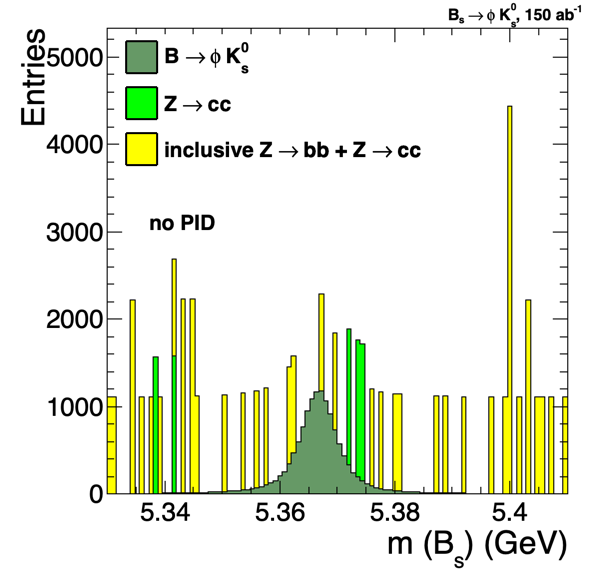}
  \end{tabular}
  \caption{\footnotesize Distribution of the mass of reconstructed $B_d$ (left) and $B_s$ (right) candidates after the final selection cuts. The plots in the top row assume perfect charged hadron PID. The lower row of plots show the distribution obtained when no PID requirement is applied to the $\phi$ and $K_s$ tracks.  
  }
  \label{fig:masses_final}
\end{figure}

\noindent The final selection cuts are summarised in Tab.~\ref{tab:full_selection}.
For simplicity, the same cuts (apart from the mass window) are used for selecting the $B_d$ and the $B_s$ signals (for the $B_d$, the aforementioned kinematic cuts further suppress the $c\bar{c}$ background). They result in an efficiency of $40 \%$ for both the $B_d$ and $B_s$ decays. The mass distributions obtained with this final selection are shown in the top row of Fig.~\ref{fig:masses_final}. For the $B_s$ selection, the $b \bar{b}$ background event that is observed right on the peak corresponds to a non-resonant decay $B_s \rightarrow K^+ K^- K_s$. Assuming that the mass distribution of the background is flat in the range depicted in the plot, about $1100$ ($400$) $b \bar{b}$ ($c \bar{c}$) background events are expected under the $B_s$ peak, with a $50 \%$ ($100 \%$) uncertainty, for $13400$ signal candidates. Hence, despite the limited Monte-Carlo statistics, one can set a lower limit of about $5$ on the signal-to-background ratio under the peak.
Finally, the lower row of plots in Fig.~\ref{fig:masses_final} shows the distributions obtained when the PID requirement is not applied. One sees that PID capabilities are mandatory in order to extract the small $B_s$ signal with a good signal-to-background ratio: without any PID, the background under the $B_s$ peak would be as large as the signal.

\section{Sensitivity to CP parameters} 
We have generated sets of events corresponding to the figures in Table~\ref{tab:stat} with their time dependence as written in equations~(\ref{eq:time}) using the parameters in Table~\ref{tab:QCDF}. The value of $\alpha_{ds}$ was set to 0.4 rad, which is close to the expected value from the SM.

\vskip 10pt
\noindent In order to study the time dependence, one needs to identify the nature of the initial B meson (B or $\rm \overline{B}$) decaying to $\rm \phi K^{*0}\overline{(K}^{*0})$ at $t=0$. This is done by tagging. The useful observed events are the tagged ones. Their time-dependent decay rate reads  
\begin{equation}
\begin{array}{ccccl} 
\Gamma(\overline{B}_{tagged}(t)\to f) & = &  (1-\omega) \Gamma(\overline{B}_q(t)\to f) + \omega \Gamma({B}_q(t)\to f) \\
\Gamma({B}_{tagged}(t)\to f) & = &  (1-\omega) \Gamma({B}_q(t)\to f) + \omega \Gamma(\overline{B}_q(t)\to f) \\
\end{array}
\label{eq:tagging}
\end{equation}

\noindent where $\omega$ is the fraction of wrong tagging. It is thus important to determine $\omega$, since it damps the amplitude of the oscillations. The quality of tagging is quantified by the figure of merit $\epsilon (1-2\omega)^2$ shown in Table~\ref{tab:Bs_tagging}.

\begin{figure}[hbt]
\vfill
\begin{center}

\includegraphics[width=0.80\textwidth]{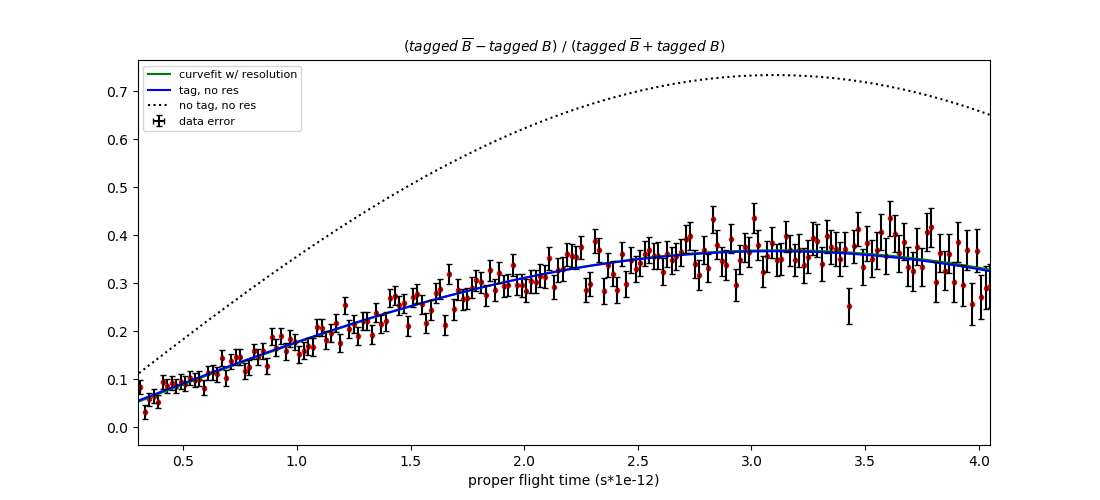}
\includegraphics[width=0.80\textwidth]{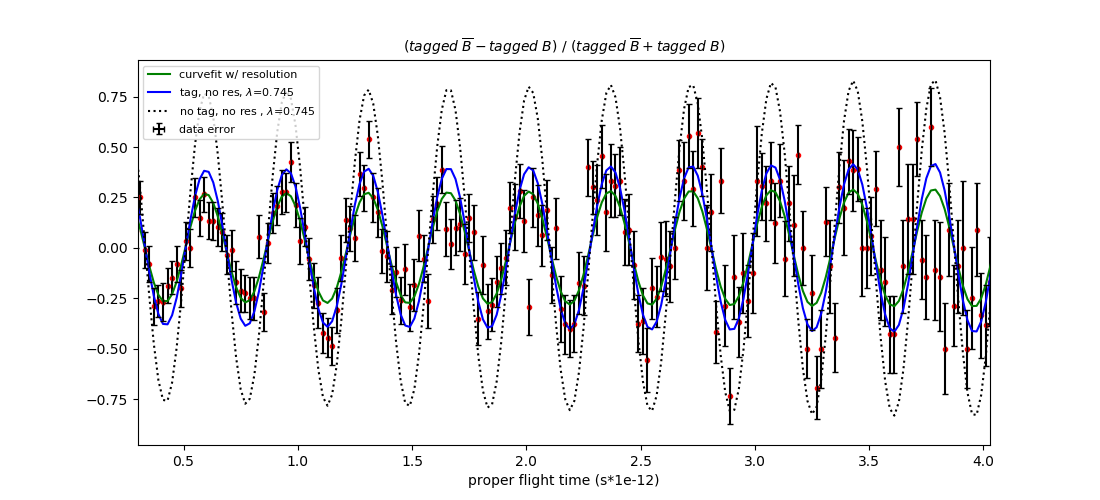}

\caption{\footnotesize \label{fig:asym_Bs} The  $B_d$ (top) and $B_s$ (bottom) CP asymmetry for the events $\overline{B}_{d,s}\to \phi K_s$. The red points with error bars are the signal. The dotted line shows the asymmetry with $\omega=0.$  and no vertex resolution. The blue line corresponds to the asymmetry with $\omega = 0.25$ and  no vertex resolution. The green line is the fit to the data.}
\end{center}
\vfill

\end{figure}

\begin{table}[htb]
\centering
$$ \begin{tabular}{cccc}
\hline
Tagging Merit & LEP & BaBar & LHCb  \\
\hline\hline
$\epsilon(1-2\omega)^2$ & 25-30\% & ~30\% & ~6\%\\
\hline

\end{tabular}   $$

\caption{\footnotesize \label{tab:Bs_tagging} Typical tagging Figure of Merit for some experiments. $\epsilon$ is the tagging efficiency and $\omega$, the wrong tagging fraction, which is in the range $0-0.5$.}
\end{table}
\vskip 10pt
\noindent Let us define the time-dependent asymmetry :
\begin{equation}
{\cal A}s = \frac{ \Gamma(\overline{B}_{tagged}(t)\to f) - \Gamma({B}_{tagged}(t)\to f) }{ \Gamma(\overline{B}_{tagged}(t)\to f)+ \Gamma({B}_{tagged}(t)\to f) }
\label{eq:asym}
\end{equation}
\noindent We show in Figure~\ref{fig:asym_Bs} the asymmetry as defined in equation~(\ref{eq:asym}).
\noindent Obviously, one needs to measure $\omega$ precisely. Fortunately, this can be done using the decay $\overline{B_s}\to D^+_s \pi^-$. The method has been described in detail in~\cite{AOP:1}. Thanks to the large statistics at FCC, an uncertainty at the sub \textperthousand\ level can be obtained for $\omega$. In the following we have assumed conservatively $\omega = 0.25$ with negligible uncertainty and a tagging efficiency of $ 100 \%$.

\begin{figure}[hbt]
\vfill
\begin{center}

\includegraphics[width=0.45\textwidth]{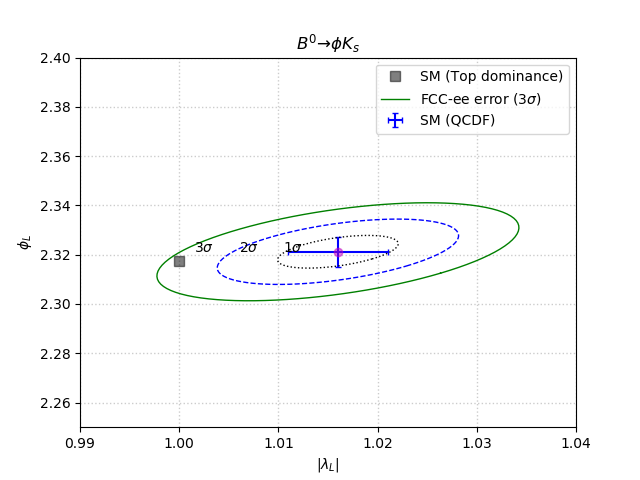}
\includegraphics[width=0.45\textwidth]{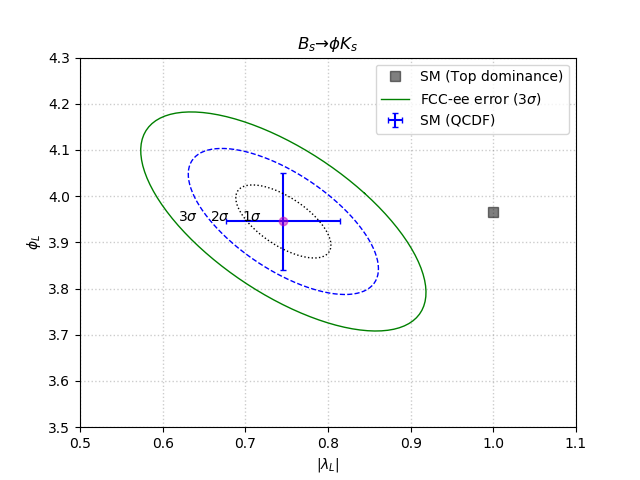}

\caption{\footnotesize \label{fig:lambda-phi_phiKs} The sensitivities for the measurements of $\lambda_L$ and $\phi_L$ (in radian) for $\overline{B}_{d,s}\to \phi K_s$.}
\end{center}
\vfill

\end{figure}

\begin{figure}[hbt]
\vfill
\begin{center}

\includegraphics[width=0.45\textwidth]{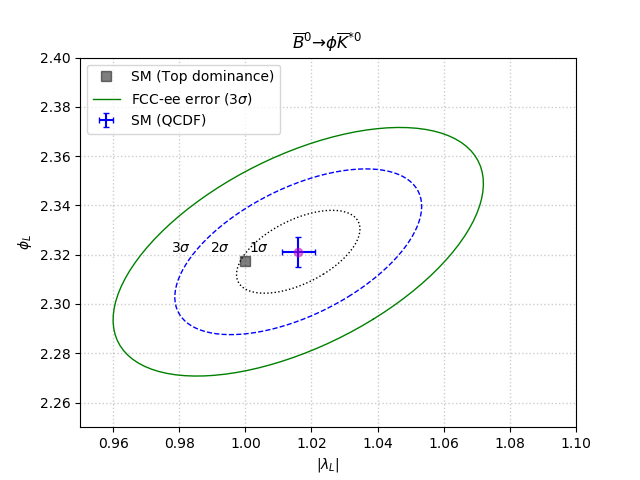}
\includegraphics[width=0.45\textwidth]{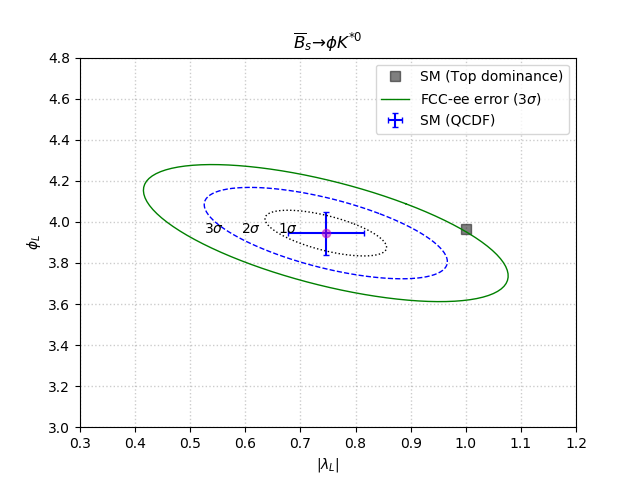}

\caption{\footnotesize \label{fig:lambda-phi_phiKstar} The sensitivities for the measurements of $\lambda_L$ and $\phi_L$ (in radian) for $\overline{B}_{d}\to \phi \overline{K}^{*0}$ and $\overline{B}_{s}\to \phi {K}^{*0}$ assuming that an amplitude analysis disentangles the various polarisation configurations.}
\end{center}
\vfill

\end{figure}

\noindent Figures~\ref{fig:lambda-phi_phiKs} and~\ref{fig:lambda-phi_phiKstar} show the results of the fits for the values of $\lambda_L$ 	and $\phi_L^{CKM}$.

\noindent The sensitivities for the modes $\overline{B}_{d}\to \phi \overline{K}^{0}$ and $\overline{B}_{s}\to \phi {K}^{0}$, which one expects at FCC-ee,  are summarized in Table~\ref{tab:sensKs}.

\begin{table}[htb]
\centering
\small
$$ \begin{tabular}{cccccc}
\hline

$\displaystyle {\mathrm {Decay}} $ &
$\displaystyle {\mathrm {\sigma(|\lambda_L|)}} $ & 
$\displaystyle {\mathrm {\sigma(\alpha_{ds})(rad)}} $ &
 \\

\hline \hline

$\displaystyle \overline{B}_d\to \phi \overline{K}^0$ &
$\displaystyle 0.005$ &
$\displaystyle {\mathrm {0.004}} $ &
\\ 

$\displaystyle \overline{B}_s\to \phi K^0$ &
$\displaystyle 0.039 $ &
$\displaystyle {\mathrm {0.045}} $ &
\\ 
\hline\hline

\end{tabular}   $$

\label{tab:sensKs}
\caption{\footnotesize \label{tab:sensKs} The expected sensitivities for $\lambda_L$ and $\alpha_{ds}$ with the decays $\overline{B}_{d}\to \phi \overline{K}^{0}$ and $\overline{B}_{s}\to \phi {K}^{0}$ .}
\end{table}

\noindent One could improve the sensitivies in Table~\ref{tab:sensKs} by using also the final states with 2 vector particles however an angular analysis should be carried out in order to take full advantage of all events. A simpler way is to deal with the events as for pseudocalar-vector decays. In that case, the CP asymmetry is further damped by the factor $\eta_f(f_L + f_\parallel - f_\perp)$. For $\phi K^{*0}$, $\eta_f = +1$, hence, according to Table~\ref{tab:B_decays}, the dilution factors are about 0.55 and 0.44 for $\overline{B}_{d}\to \phi \overline{K}^{*0}$ and $\overline{B}_{s}\to \phi {K}^{*0}$, respectively.
\noindent The sensitivities for the modes $\overline{B}_{d}\to \phi \overline{K}^{*0}$ and $\overline{B}_{s}\to \phi {K}^{*0}$, which one expects at FCC-ee (see Table~\ref{tab:sensKs*}),  are significantly worse than for  ${\rm \phi {K}^{0}}$ because only ${\rm K^{*0} \to K^0 \pi^0}$ decays can be used. A more complete analysis using the angular dependences of the polarization states would enable to improve the sensitivities by about a factor of 2 as displayed in the Figure~\ref{fig:lambda-phi_phiKstar}.

\begin{table}[htb]
\centering
\small
$$ \begin{tabular}{cccccc}
\hline

$\displaystyle {\mathrm {Decay}} $ &
$\displaystyle {\mathrm {\sigma(\lambda_L)}} $ & 
$\displaystyle {\mathrm {\sigma(\alpha_{ds})(rad)}} $ &
 \\

\hline \hline

$\displaystyle \overline{B}_d\to \phi \overline{K}^{*0}$ &
$\displaystyle 0.012$ &
$\displaystyle {\mathrm {0.022}} $ &
\\ 

$\displaystyle \overline{B}_s\to \phi K^{*0}$ &
$\displaystyle 0.07 $ &
$\displaystyle {\mathrm {0.14}} $ &
\\ 
\hline\hline

\end{tabular}   $$

\label{tab:sensKs*}
\caption{\footnotesize \label{tab:sensKs*} The expected sensitivities for $\lambda_L$ and $\alpha_{ds}$ with the decays $\overline{B}_{d}\to \phi \overline{K}^{*0}$ and $\overline{B}_{s}\to \phi {K}^{*0}$ .}
\end{table}
\noindent  Finally, it is very important to note that these modes require an outstanding electromagnetic calorimeter in order to get a manageable background both due to the combinatorics and, in the case  $\overline{B}_{s}\to \phi{K}^{(*)0}$, from the decay $\overline{B}_{d}\to \phi \overline{K}^{(*)0}$, which would contaminate the former decay, should the photon resolution not be very good, more quantitatively $dE/E \leq 0.03/\sqrt{E({\rm GeV})} +0.003$ is necessary.
\section{Conclusions}
\noindent We have shown that it is possible to measure one of the angles of the flattest unitarity triangle, namely $\alpha_{ds}$, using the decays $\overline{B}_{d}\to \phi \overline{K}^{(*)0}$ and $\overline{B}_{s}\to \phi{K}^{(*)0}$. Very interesting sensitivities, better than $ 4 {\mbox{mrad}}$, are expected at FCC-ee with an integrated luminosity of 150 $ab^{-1}$ at a center of mass energy $E_{cm}=M_Z$ allowing one to perform further tests of the Standard Model. This measurement requires an excellent tracking system. In particular, a large tracking volume with many measurement layers is crucial for reconstructing $K_s$ decays up to large flight distances, and a light tracker and a highly performant vertex detector are needed. Moreover, extracting the small $B_s$ signal from the background requires charged hadron PID, and reconstructing the modes with a $K^{*0}$ demands an outstanding resolution of the electromagnetic calorimeter.
{\subsection*{Acknowledgments}}
We wish to thank Franco Bedeschi for making his vertexing code available and for very useful discussions about the reconstruction of displaced vertices. \\
\par


\vskip 10pt
\newpage
\begin{thebibliography}{99}
\bibitem{fccee:1}M. Bicer, et al., {\it  First look at the physics case of TLEP}, J. High Energy Phys. 01 (2014) 164, https://doi .org /10 .1007 /JHEP01(2014 )164, arXiv:1308 .6176.
\bibitem{fccee:2} A. Abada, et al., FCC Collaboration, {\it FCC Physics Opportunities : Future Circular Collider Conceptual Design Report Volume 1}, Eur. Phys. J. C 79(6) (2019) 474, https://doi .org /10 .1140 /epjc /s10052 -019 -6904 -3.
\bibitem{fccee:3} A. Abada, et al., FCC Collaboration, {\it FCC-ee: The Lepton Collider : Future Circular Collider Conceptual Design Report Volume 2}, Eur. Phys. J. ST 228(2) (2019) 261, https://doi .org /10 .1140 /epjst /e2019 -900045 -4.

\bibitem{AOP:1} R. Aleksan, L. Oliver and E. Perez, {\it CP violation and determination of the bs ``flat'' unitarity triangle at FCCee} \Journal\PRD&105&(2022) 5, 053008  [arXiv:2107.02002[hep-ph]].
\bibitem{AOP:2} R. Aleksan, L. Oliver and E. Perez, {\it Study of CP violation in $B^\pm$ decays to $\overline{D^0}(D^0)K^\pm$ at FCCee}, [arXiv:2107.05311[hep-ph]].
\bibitem{CKM:1} M. Kobayashi, T. Maskawa, {\it CP-Violation in the Renormalizable Theory of Weak Interaction} \Journal\PTP&49&(1973) 652.
\bibitem{akl:1} R. Aleksan, B. Kayser and D. London, {\it Determining the Quark Mixing Matrix From CP-Violating Asymmetries} \Journal\PRL&73&(1994) 18-20 , arXiv:hep-ph/9403341.
\bibitem{pdg:1} P.A. Zyla \etal\ [Particle Data Group] \Journal\PTEP&&(2020) 083C01.
\bibitem{AO:1} R. Aleksan and L. Oliver, in preparation.
\bibitem{de_Favereau_2014} J. de Favereau, C. Delaere, P. Demin, A. Giammanco, V. Lema$\hat{\rm \i}$tre, A. Mertens and M. Selvaggi, {\it Delphes 3: a modular framework for fast simulation of a generic collider experiment}, JHEP 2014 (Feb, 2014).
\bibitem{BedeschiCode} F. Bedeschi. Code available as part of the TrackCovariance module of the Delphes package, https://github.com/delphes/delphes.
\bibitem{Franco_neutrals} F. Bedeschi. Presentation at the FCC Physics Performance meeting, October 2023, https://indico.cern.ch/event/1337943/.
\bibitem {Bierlich:2022pfr} C. Bierlich et al., {\it A comprehensive guide to the physics and usage of PYTHIA 8.3}, SciPost Phys. Codeb. 2022 (2022) 8, [arXiv:2203.11601[hep-ph]].
\bibitem {LANGE2001152}  D. J. Lange, {\it The EvtGen particle decay simulation package}, Nucl. Instrum. Meth. A {\bf 462} (2001).
\bibitem{FCCAnalyses} C. Helsens and the FCC software group, https://github.com/HEP-FCC/FCCAnalyses.
\end{thebibliography}
\end{document}